\documentclass[aps,prc,preprint,showpacs,preprintnumbers,amsmath,amssymb] {revtex4-1}
\usepackage{amsmath}
\usepackage{amssymb}
\usepackage{graphicx}
\usepackage{subfigure}
\usepackage{rotating}
\usepackage{natbib}
\usepackage{txfonts}
\usepackage{color}
\usepackage{bbold}
\def\beq{\begin{equation}}
\def\eeq{\end{equation}}
\def\bea{\begin{eqnarray}}
\def\eea{\end{eqnarray}}

\def\eqref#1{Eq.~(\ref{eq:#1})}

\def\be{\begin{equation}}
\def\ee{\end{equation}}
\def\bg{\begin{eqnarray}}
\def\en{\end{eqnarray}}

\long\def\Omit#1{}

\DeclareGraphicsExtensions{ .pdf, .jpg, .eps}
\begin{document}
\title{Reaction ${\bar p}p \to {\bar \Lambda}_c^-\Lambda_c^+$ within an 
effective Lagrangian model}
\author{R. Shyam$^1$}
\author{H. Lenske$^2$}
\affiliation{$^1$Saha Institute of Nuclear Physics, 1/AF Bidhan Nagar, Kolkata 700064, 
India and Physics Department, Indian Institute of Technology, Roorkee 247667, India{
\footnote {E-mail: radhey.shyam@saha.ac.in}}}
\affiliation{$^2$Institut f\"ur Theoretische Physik, Universit\"at Gieseen, 
Heinrich-Buff-Ring 16, D-35392 Giessen, Germany} 

\date{\today}
\begin{abstract}
We study the charmed baryon production reaction ${\bar p} p \to {\bar \Lambda}_c^- 
\Lambda_c^+$ using the $t$-channel $D^0$ and $D^{*0}$ meson-exchange diagrams 
within an effective Lagrangian model involving the physical hadron masses and the 
coupling constants determined from SU(4) flavor symmetry. The initial and final 
state distortion effects are accounted for by using a simple eikonal 
approximation-based procedure. The vertex parameters of our model have been checked 
by employing them to calculate the cross sections for the ${\bar p} p \to {\bar 
\Lambda} \Lambda$ reaction within a similar model. We predict the ${\bar 
\Lambda}_c^- \Lambda_c^+$ production cross sections in the range of 1$-$30 $\mu b$ 
for antiproton beam momenta varying between threshold and 20 GeV/$c$. The respective 
roles of $D^0$ and $D^{*0}$ meson exchanges and also those of the vector and tensor 
components of the $D^{*0}$ coupling have been investigated.
\end{abstract}
\pacs{13.75.Cs, 14.20.Lq, 11.10.Ef}
\maketitle

\section{Introduction}

The heavy hadrons consisting of a charm quark are quite distinct in their 
properties from the light flavored hadrons composed of up ($u$), down ($d$), and 
strange ($s$) quarks. The presence of the heavy quark in heavy flavor hadrons 
provides an additional handle for the understanding of quantum chromodynamics (QCD), 
the fundamental theory of the strong interaction. The large mass of the charm quark 
introduces a mass scale much larger than the confinement scale $\Lambda_Q \approx 
300$ MeV. In contrast, the energy scale of the lighter quarks is $\ll \Lambda_Q$. 
The presence of two scales in such systems naturally leads to the construction of 
an effective theory where one can actually calculate a big portion of the relevant 
physics using perturbation theory and renormalization-group techniques. The heavy 
quark effective theory (HQET) is one such approach. New symmetry properties, not 
apparent in QCD, appear in HQET~\cite{isg89,kor94,neu94,gro99,man00}. The charm 
hadrons are ideal candidates to test and apply the predictions of HQET.
 
In this context, the investigation of the production of heavy flavor hadrons 
is of great interest.  Since the discovery of $J/\psi$ in 1974~\cite{aub74,aug74},  
the production of charmonium ($c{\bar c}$) states has been extensively studied  
experimentally in hadroproduction (Tevatron)(see, eg. the reviews~\cite{swa06,god08}),
photo- and electro-production (HERA) (see, e.g., Refs.~\cite{kat03,ste08} for 
details) and $e^-e^+$ annihilation ($BABAR$, Belle, and BES) (see, e.g.,  Refs.
\cite{san13,uch13,pin13} for recent reviews) reactions. There are also a large 
number of theoretical studies of the charmonium production (see, e.g., Ref.
\cite{bod12} for a review). These studies have contributed substantially to 
enhance our understanding of the charm meson states, their spectroscopy, and decays.
 
The first charmed baryon states were detected in 1975 in neutrino interactions
\cite{cas75}. Since then, many new excited charmed baryon states have been 
discovered by the CLEO~\cite{dub03}, $BABAR$~\cite{zie11}, and Belle~\cite{kat14} 
facilities (Ref.~\cite{kor94} provides a good review of the older studies). However, 
the production and spectroscopy of the charmed baryons have not been explored in 
the same detail as the charmonium states, although they can provide similar 
information about the quark confinement mechanism. In fact, due to the presence of
three quarks (two light and one heavy), the structure of the charmed baryon is more
intriguing and complicated. In contrast to mesons, there can be more states as 
there are more possibilities of orbital excitations.

Most of the current experimental information about the production of the ground
state charmed baryon [$\Lambda_c(2286)$] has been derived from the 
electron-positron annihilation experiments. In the near future, charmed baryon 
production will be studied in the proton-antiproton ($p{\bar p}$) annihilation 
using the "antiproton annihilation at Darmdtadt" (${\bar P}ANDA$) experiment at 
the Facility for Antiproton and Ion Research (FAIR) in GSI, Darmstadt (see, e.g., 
Ref.~\cite{wie11}). The advantage of using antiprotons in the study of the 
charmed baryon is that in $p{\bar p}$ collisions the production of extra particles 
is not needed for charm conservation, which reduces the threshold energy as 
compared to, say, $pp$ collisions. The beam momenta of antiprotons in this experiment 
will be well above the threshold (10.162 GeV$/ c$) of the ${\bar p} p \to 
{\bar \Lambda}_c^- \Lambda_c^+$ reaction. For the planning of this experiment, 
reliable theoretical estimates of the cross section of this reaction would be of 
crucial importance. 

Several authors have calculated the cross section of this reaction by using
a variety of models~\cite{kro89,kai94,tit08,gor09,hai10,hai11,kho12}. However, 
the magnitudes of the predicted cross sections are strongly model dependent $-$
they differ from each other by several orders of magnitudes. Furthermore, there 
is no unanimity about the degrees of freedom to be used in order to describe 
this reaction. In Ref.~\cite{gor09}, the ${\bar p} p \to {\bar \Lambda}_c^- 
\Lambda_c^+$ reaction has been described within a handbag approach where the 
amplitude is calculated by convolutions of hard subprocess kernels (representing 
the process $u{\bar u} \to c{\bar c}$ ) and the generalized parton distributions, 
which represent the soft nonperturbative physics. This approach bears some 
resemblance to the quark-diquark picture used by some of these authors in 
Ref.~\cite{kro89} to make predictions for the cross sections of the 
${\bar \Lambda}_c^- \Lambda_c^+$ production. In the study reported in 
Ref.~\cite{kai94}, a quark-gluon string model together with Regge asymptotics 
for hadron amplitudes has been used. Calculations reported in 
Refs.~\cite{tit08,kho12} are also based on similar ideas. 

On the other hand, in Refs.~\cite{hai10,hai11}, a meson-exchange model was 
used to describe the ${\bar p} p \to {\bar \Lambda}_c^- \Lambda_c^+$ reaction.
This approach is based on the J\"ulich meson-baryon model that was employed 
earlier~\cite{hai92a,hai92b} to investigate the ${\bar p} p \to {\bar \Lambda} 
\Lambda$ reaction. In this model, these reactions are considered within a 
coupled-channels framework, which allows one to take into account the initial 
and final state interactions in a rigorous way. The reaction proceeds via an 
exchange of appropriate mesons between $p$ and ${\bar p}$ leading to the
final baryon-antibaryon state. Also in Ref.~\cite{he11} the meson-exchange 
picture was used to calculate the production rate of the charmed baryon
$\Lambda_c^+(2940)$ in the $p{\bar p}$ annihilation at ${\bar P}ANDA$ energies.

The aim of this paper is to investigate the ${\bar p} p \to {\bar \Lambda}_c^- 
\Lambda_c^+$ reaction within a single-channel effective Lagrangian model (see, 
e.g., Refs.~\cite{shy99,shy02}), where this reaction is described as a sum of 
$t$-channel $D^0$ and $D^{*0}$ meson-exchange diagrams (see Fig.~1). The 
$\Lambda_c^+$  mass ($m_{\Lambda_c^+}$) is taken to be 2.286 GeV. The  $s$- and  
$u$-channel resonance excitation diagrams are suppressed, as no resonance with 
energy in excess of 3.0 GeV having branching ratios for decay to the 
$\Lambda_c^+$ channel is known. Although, in some chiral coupled-channel 
studies~\cite{lut05,lut06} the existence of narrow cryptoexotic baryon resonances 
with hidden charm has been predicted, these are confined in the mass range 
between 3 and 4 GeV/$c^2$ and are unlikely to contribute to the open 
charmed baryon production reaction. At the same time, the direct $p{\bar p}$ 
annihilation into ${\bar \Lambda}_c^- \Lambda_c^+ $ via the contact diagrams is 
also suppressed due to the Okubo-Zweig-Iizuka condition.

In the next section, we present our formalism. The results and discussions of our
work are given in Sec. III. Finally, the summary and the conclusions of this study
are presented in Sec. IV.  
 
\section{Formalism}

\begin{figure}[t]
\centering
\includegraphics[width=.50\textwidth]{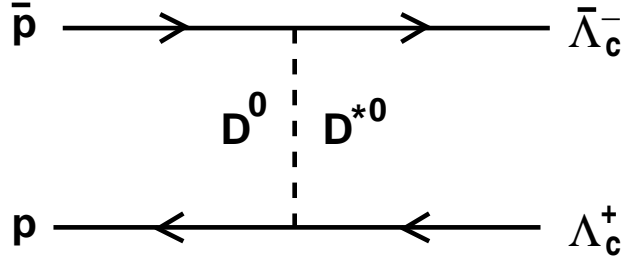}
\caption{(color online)
Graphical representation of the model used to describe the ${\bar p} + p \to
{\bar \Lambda}_c^- + \Lambda_c^+$ reaction. $D^0$ and $D^{*0}$ in the 
intermediate line represent the exchanges of $D^0$ pseudoscalar and $D^{*0}$
vector mesons, respectively. 
}
\label{fig:Fig1}
\end{figure}
To evaluate various amplitudes for the processes shown in Fig.~1, we have used the
effective Lagrangians at the charm baryon-meson-nucleon vertices, which are taken 
from Refs.~\cite{hob13,hai11a,gro90}. For the $D^0$ meson [mass ($m_{D^0}$)
= 1.865 GeV] exchange vertices we have 
\begin{eqnarray}
{\cal L}_{D^0BN} & = & ig_{BD^0N}{\bar \psi}_B i\gamma_5 \psi_N \phi_{D^0} + H.c.,
\end{eqnarray}
where ${\psi}_B$ and $\psi_N$ are the charmed baryon and nucleon (antinucleon)
fields, respectively, and $\phi_{D^0}$ is the $D^0$ meson field. $g_{BD^0N}$
in Eq.~(1) represents the vertex coupling constant.

For the $D^{*0}$ meson [mass ($m_{D^{*0}}$) = 2.007 GeV] exchange vertices, the 
effective Lagrangian is
\begin{eqnarray}
{\cal L}_{D^{*0}BN} & = & g_{D^{*0}BN} {\bar \psi}_B \gamma_\mu \psi_N 
                         \theta_{D_{*0}}^\mu + \frac{f_{D^{*0}BN}}{4M}{\bar \psi}_B
                         \sigma_{\mu \nu} \psi_N F_{D^{*0}}^{\mu \nu} + H.c.,
\end{eqnarray}
where $\theta_{D_{*0}}^\mu$ is the vector meson field, with field strength tensor
$F_{D^{*0}}^{\mu \nu} = \partial^\mu \theta_{D^{*0}}^\nu - 
\partial^\nu \theta_{D^{*0}}^\mu$. $\sigma_{\mu \nu}$ is the usual tensor operator.
The vector and tensor couplings are defined by $g$ and $f$, respectively, which 
were fixed in Refs.~\cite{hai07,hai08,hai11a} by using SU(4) symmetry arguments in the 
description of the exclusive charmed hadron production in ${\bar D}N$ and $DN$ 
scattering within a  one-boson-exchange picture. In our study we have adopted the 
values as given in Ref.~\cite{hai11a} (see Table I). The same couplings were used 
for the vertices involving both the proton and the antiproton. It may be pointed out 
here that in the study presented in Ref.~\cite{he11}, the exchange of $D^{*0}$ was 
not considered. As we shall show later on, this process dominates the $\Lambda_c^+ 
{\bar \Lambda}_c^-$ production reaction in the $p{\bar p}$ annihilation even for 
beam momenta closer to the production threshold. 

\begin{table}[here]
\begin{center}
\caption {Coupling constants at the $BD^0N$ and $BD^{*0}N$ vertices. These are 
taken from Ref.~\cite{hai11a} where they are deduced from $DN$ and ${\bar D}N$ 
scattering analysis. Here $B$ represents the charmed baryon.
}
\vspace{0.5cm}
\begin{tabular}{|c|c|c|}
\hline
Vertex  & $g_{DBN}/\sqrt{4\pi}$ & $f_{DBN}/\sqrt{4\pi}$ \\
\hline
$ND^0B$     & 3.943 & --   \\
$ND^{*0}B$  & 1.590 & 5.183\\
\hline
\end{tabular}
\end{center}
\end{table}

The off-shell behavior of the vertices is regulated by a monopole form factor
(see, eg., Refs.~\cite{shy99,shy02})
\begin{eqnarray}
F_i(q_i) & = & \frac{\lambda_i^2-m_{D_i}^2}{\lambda_i^2-q_{D_i}^2},
\end{eqnarray}
where $q_{D_i}$ is the momentum of the {\it i}th exchanged meson with mass 
$m_{D_i}$. $\lambda_i$ is the corresponding cutoff parameter, which governs 
the range of suppression of the contributions of high momenta carried out via 
the form factor. We chose a value of 3.0 GeV for $\lambda_i$ at both the vertices. 
The same $\lambda_i$ was also used in the monopole form factor employed in the 
studies presented in Refs.~\cite{hai10,hai11}. It is  of interest to note that a 
value of $\lambda = (2.89 \pm 0.04)$ was determined in Ref.~\cite{hob13} by a 
one-boson-exchange model fitting of the inclusive $\Lambda_c^+$ production cross 
section in the proton-proton collision measured by the R680 Collaboration at ISR
\cite{cha87}.  Because the experimental data are not yet available for the reaction 
under investigation in this paper, we restrict ourselves to the choice of the form 
factor given by Eq.~(3) with a $\lambda_i$ value as mentioned above. This enables a 
meaningful comparison of our results with those of Refs.~\cite{hai10,hai11}. 

For calculating the amplitudes, we require the propagators for the exchanged mesons.
For the $D^0$ and $D^{*0}$ mesons, the propagators are given by  
\begin{eqnarray}
G_{D^0}(q) & = & \frac{i} {{q^2 - m_{D^0}^2 }},\\
G_{D^{*0}}^{\mu\nu}(q) & = & -i\left(\frac{{g^{\mu\nu}-q^\mu q^\nu/q^2}}
                           {{q^2 - (m_{D^{*0}}-i\Gamma_{D^{*0}}/2)^2}} \right).
\end{eqnarray}
In Eq.~(5), $\Gamma_{D^{*0}}$ is the total width of the $D^{*0}(2007)$ meson which 
is about 2.0 MeV according to the latest particle data group estimate~\cite{ber12}. 
After having established the effective Lagrangians, coupling constants, and 
forms of the propagators, the amplitudes of various diagrams can be written by 
following the well-known Feynman rules. The signs of these amplitudes are fixed 
by those of the effective Lagrangians, the coupling constants, and the propagators 
as described above. These signs are not allowed to change anywhere in the 
calculations.
 
From the studies of the ${\bar p} p \to {\bar \Lambda}\Lambda$ reaction, it 
is known that there is a strong sensitivity of the calculated cross sections 
to the distortion effects in the initial and final states
\cite{gen84,tab85,kro87,bur88,koh86,bri89,rob91,tab91,hai92a,hai92b,alb93}. For 
the ${\bar \Lambda}_c^- \Lambda_c^+$ production channel also the magnitudes of the 
cross sections have been found \cite{hai10} to depend very sensitively on the 
distortion effects.

For the ${\bar p} p$ initial state, the annihilation channel is almost as strong 
as the elastic scattering. This large depletion of the flux can be accounted for 
by introducing absorptive potentials that are used in optical model or in
coupled-channels approaches~\cite{koh86,hai92a,hai92b,alb93,hai10}. In this work
we do not employ such a detailed treatment. Instead, we use a procedure that 
was originated by Sopkovich~\cite{sop62}. In this method, the transition amplitude 
with distortion effects is written as
\begin{eqnarray}
T^{{\bar p}p \to {\bar \Lambda}_c^- \Lambda_c^+} & = &
\sqrt{\Omega^{{\bar p}p}}T^{{\bar p}p \to {\bar \Lambda}_c^- \Lambda_c^+}_{Born}
\sqrt{\Omega^{{\bar \Lambda}_c^- \Lambda_c^+}} 
\end{eqnarray}
where $T^{{\bar p}p \to {\bar \Lambda}_c^- \Lambda_c^+}_{Born}$ is the transition
matrix calculated within the Born approximation and $\Omega^{{\bar p}p}$ and the 
$\Omega^{{\bar \Lambda}_c^- \Lambda_c^+}$ are the matrices describing the initial 
and final state elastic scattering, respectively. Their effect is to dampen the 
wave functions and hence the amplitudes. We, however, note that the derivation of 
this equation relies on the ideas of the high-energy eikonal model while we are 
dealing with low energies in the final channels. Nevertheless, it has been shown 
in Ref.~\cite{hai10} that, because of the strong absorption in the initial channel, 
the results turn out to be rather insensitive to the final state ${\bar \Lambda}_c^- 
\Lambda_c^+$ interactions. In fact, even if the final state interactions (FSIs) are 
ignored totally, the total cross sections do not change by more than 10$\%$-15$\%$. 
In order to keep the number of free parameters small, we, therefore, decided to 
fully neglect FSIs and concentrate only on the initial state interaction.

For the present purpose, we neglect the real part of the baryon-antibaryon
interaction. Considering the ${\bar p}p$ initial state interaction, we describe the 
strong absorption by an imaginary potential of Gaussian shape with range parameter 
$\mu$ and strength $V_0$. By using the eikonal approximation, the corresponding 
attenuation integral can be evaluated in a closed form. Similar to 
Refs.~\cite{sop62,rob91}, we obtain for $\Omega^{{\bar p}p}$  
\begin{eqnarray}
\Omega^{{\bar p}p} & = & exp\big[\frac{-\sqrt{\pi}EV_0}{\mu k} exp(-\mu^2 b^2)\big],
\end{eqnarray} 
where $b$ is the impact parameter of the ${\bar p}p$ collision. $E$ and $k$ are the 
center of mass energy and the wave vector of the particular channel, respectively.
In our numerical calculations, we have used $V_0 = 0.8965$ GeV and $\mu = 0.3369$ 
GeV. With these values, the total cross sections are reasonably well described in 
the relevant energy region. For the impact parameter, we have taken a value of 
0.327 GeV$^{-1}$. With these parameters, we are able to get cross sections for the 
${\bar p}p \to {\bar \Lambda}\Lambda$ reaction in close agreement with the 
corresponding experimental data. Furthermore, they lead to cross sections for 
the ${\bar \Lambda}_c^- \Lambda_c^+$ production that are similar in magnitude to 
those reported in Ref.~\cite{hai10}.

Although the parameters $V_0$ and $\mu$ may change with energy, we have made them 
global; that is, they remain the same at all the energies. Furthermore, the same 
parameters were used in the calculations of both the ${\bar p} p \to {\bar \Lambda}
\Lambda$ and the ${\bar p}p \to {\bar \Lambda}_c^- \Lambda_c^+$ reactions. Thus,
we have only three fixed parameters in our calculations of the initial state 
distortion effects.

\section{Results and Discussions}

\begin{figure}[t]
\centering
\includegraphics[width=.50\textwidth]{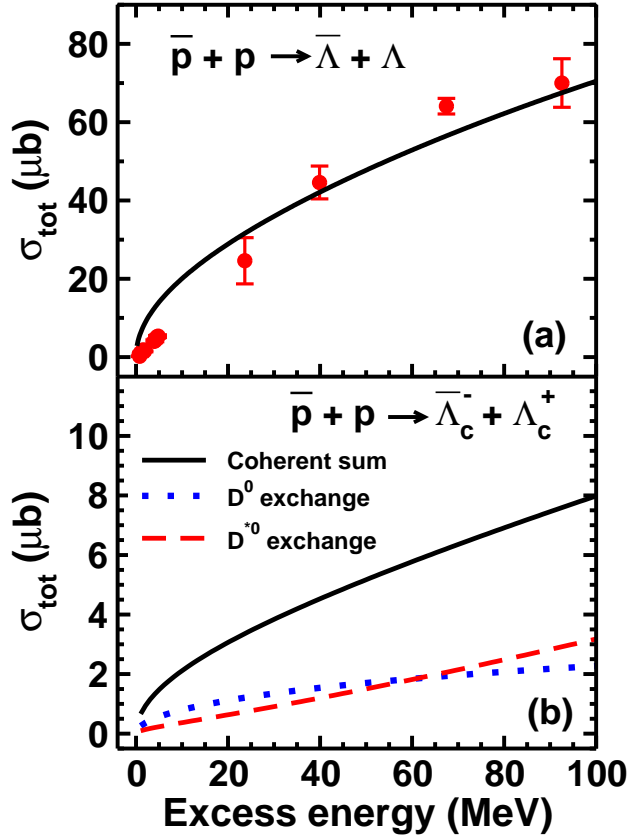}
\caption{(color online)
Total cross section for the reactions ${\bar p} p \to {\bar \Lambda} \Lambda$ (a) 
and ${\bar p}p \to {\bar \Lambda}_c^- \Lambda_c^+$ (b) as a function of the excess 
energy.  The experimental data in (a) are taken from ~\protect\cite{bar89,bar91,bar96}. 
In (b) the contributions of the $D^0$ and $D^{*0}$ exchange processes are shown by 
dotted and dashed lines, respectively. The solid line represents their coherent sum.
}
\label{fig:Fig2}
\end{figure}
The formalism described above has been used first to describe the ${\bar p} p \to 
{\bar \Lambda} \Lambda$ reaction where experimental data are available for both
near-threshold and far-from-threshold beam momenta. The aim is to check the 
parameters of our model (the coupling constants and those related to the distortion
effects). In Fig. 2(a), we show the total cross section of this reaction for beam
momenta closer to the reaction threshold (1.433 GeV/$c$) as a function of the excess 
energy (defined as $\sqrt{s}-m_{\bar \Lambda}-m_{\Lambda}$, with $\sqrt{s}$ being 
the invariant mass). In these calculations, we have considered the exchange of 
pseudoscalar $K(498)$ and pseudovector $K^*(892)$ mesons. A width of 48 MeV is 
taken in the denominator of the $K^*(892)$ propagator [Eq.~(5)]. The effective 
Lagrangians for the baryon-meson-nucleon vertices were the same as those given by 
Eqs.~(1) and (2). Assuming a complete SU(4) symmetry, the values of the vertex 
coupling constants were taken to be the same as those shown in Table 1. The 
parameters of the initial state distortion factor were also  the same as described 
above. However, as in Refs.~\cite{hai92a,hai92b}, the cutoff parameter of the form 
factor in Eq.~(3) was chosen to be 1.7 GeV due to the different mass regime of the 
exchanged mesons. The experimental data in Fig.~2(a) are taken from  
Refs.~\cite{bar89,bar91,bar96}. We note that there is a good overall agreement with 
the data for excess energies ranging between threshold and 100 MeV. 

In Fig.~2(b), our results for the ${\bar p}p \to {\bar \Lambda}_c^- \Lambda_c^+$ 
reaction are shown as a function of the corresponding excess energy (defined in this
case as $\sqrt{s}-m_{{\bar \Lambda}_c}^--m_{{\Lambda}_c}^+$). Because of the assumption 
of SU(4) symmetry, all the coupling constants were taken to be the same. Even the 
parameters involved in the ${\bar p}p$ distortion factors were the same. However, 
the cutoff parameter $\lambda$ in this case was 3 GeV as discussed in the previous 
section. We notice that the magnitude of the cross section in Fig.~2(b) is smaller 
than that of Fig.~2(a) by nearly an order of magnitude. This is in agreement with the 
results obtained in the coupled-channel calculations presented in Ref.~\cite{hai10}.
This difference has been attributed to the difference in the masses of the mesons 
involved in the corresponding propagators.  We further note that very close to the 
threshold the contributions of the $D^0$ exchange are slightly larger than those 
of the $D^{*0}$ exchange. However, with the increasing beam momentum, the situation 
reverses, and the $D^{*0}$ exchange starts dominating the total cross section. The 
absolute magnitudes of the cross sections in Fig. 2(b) are very similar to those of 
Ref.~\cite{hai10}. 

\begin{figure}[t]
\centering
\includegraphics[width=.50\textwidth]{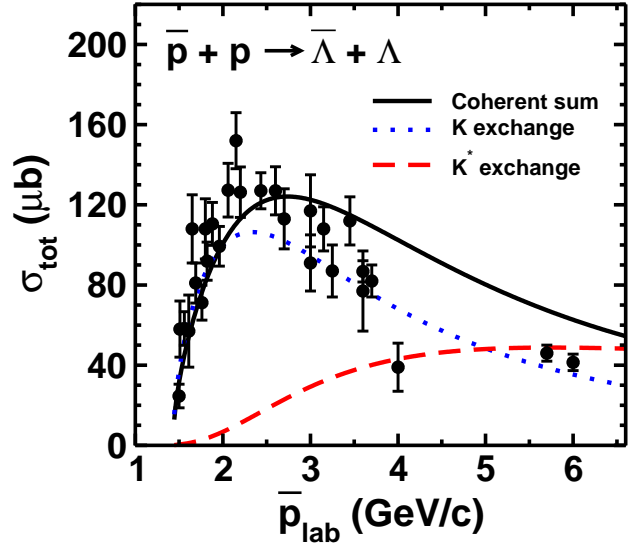}
\caption{(color online)
Total cross section for the reaction ${\bar p}p \to {\bar \Lambda} \Lambda$ 
as a function of the antiproton beam momentum. The individual contributions of 
$K(494)$ and $K^*(892)$ exchange processes are shown by dotted and dashed lines, 
respectively. The solid line represents their coherent sum.
}
\label{fig:Fig3}
\end{figure}

We next discuss our results at higher beam momenta. In Fig.~3, we compare the 
calculated total cross sections for the ${\bar p} p \to {\bar \Lambda} \Lambda$ 
with the corresponding experimental data that are available for incident ${\bar p}$ 
beam momenta (${\bar p}_{lab}$) from near threshold to above 6 GeV/$c$~\cite{fla84}.
It is clear that our calculations provide a reasonable description of the 
beam momentum dependence of the data. In this figure, we also show the individual 
contributions of the $K(494)$ and $K^*(892)$ meson-exchange processes. We note that 
for beam momenta near the threshold (${\bar p}_{lab} < $ 2.0 GeV/$c$) the 
$K$-exchange terms are dominant.  However, for ${\bar p}_{lab}$ beyond this range the 
$K^*$-exchange process becomes important, and for ${\bar p}_{lab} > $ 6.0 GeV/c it 
contributes most to the total cross section. This result is in agreement with the 
observations made in several previous studies (see, e.g., Ref.~\cite{tab85} and the 
references of the older works cited there). 

Some disagreement seen in Fig.~3 between the theory and the data for 
${\bar p}_{lab}$ beyond $\geq$ 4.0 GeV/$c$ could be an indication of the increasing 
importance of larger mass strange meson contributions at higher beam momenta. It 
has been noted in Ref.~\cite{tab85} that, at higher beam energies, the $K_2^*(1430)$ 
meson-exchange process becomes crucial, as it interferes destructively with the 
$K^*(892)$ terms. In any case, since we are mainly interested in the overall 
mechanism of the production of the flavored baryon-antibaryon pair in the ${\bar p}p$ 
annihilation reaction, and also since there are large uncertainties in the data, 
we refrain from attempting a detailed fit to the data.
 
For the charm baryon production reaction, we investigate the role of various 
meson-exchange processes at higher beam momenta in Fig.~4. In this figure, we show 
the total cross sections for the ${\bar p}p \to {\bar \Lambda}_c^- \Lambda_c^+$ 
reaction for ${\bar p}_{lab}$ varying in the range of threshold to 20 GeV/$c$. First 
we note that the cross sections peak around ${\bar p}_{lab}$ of 15 GeV/$c$, and 
thereafter they decrease gradually. We further see that the vector meson ($D^{*0}$) 
exchange process dominates the cross section except for very-close-to-threshold 
beam momenta.  For ${\bar p}_{lab}$ $\geq$ 15 GeV/$c$, the $D^0$-exchange contributions 
are nearly an order of magnitude smaller than those of the $D^{*0}$-exchange. 
However, it is also clear in this figure that, even though for ${\bar p}$ beam momenta 
away from the threshold the individual contributions of the $D^0$ exchange processes
are small, they are not negligible, as they contribute significantly through the 
interference terms which are constructive for this case.
\begin{figure}[t]
\centering
\includegraphics[width=.50\textwidth]{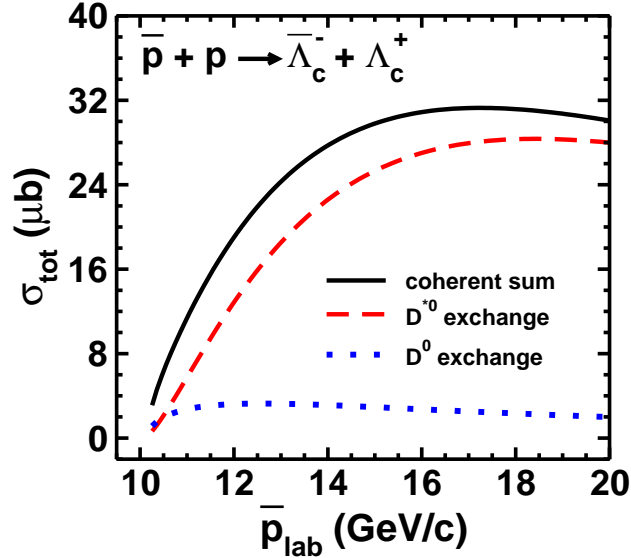}
\caption{(color online)
Total cross section for the reaction ${\bar p}p \to {\bar \Lambda}_c^- \Lambda_c^+$ 
as a function of the antiproton beam momentum. The contributions of the $D^0$ and 
$D^{*0}$ exchange processes are shown by dotted and dashed lines, respectively. 
The solid line represents their coherent sum.
}
\label{fig:Fig4}
\end{figure}

To explore the origin of the domination of the $D^{*0}$ exchange contributions, 
we note the relative strong coupling to the vector meson vertices, particularly
for the tensor coupling term. As can be seen from Table I, the ratio of tensor to 
vector coupling is 3.26. This analogous to the large tensor coupling for the 
$\rho$ meson in the one-boson-exchange models of the $NN$ interaction~\cite{mac87},
where the tensor to vector ratio is even larger (6.1). In Fig.~5, we show the 
individual contributions of the vector and tensor terms to the total $D^{*0}$ 
exchange cross section. Clearly, the tensor coupling terms make the dominant 
contribution to the $D^{*0}$ exchange part of the total cross section. The 
interference of vector and tensor coupling terms is destructive as the total 
$D^{*0}$ cross sections are lower than the individual contribution of the tensor 
term.  At the effective Lagrangian level, the strong tensor contribution is 
associated with the additional momentum dependence induced by the derivative 
coupling in the tensor interaction [see Eq.~(2)].

A comparison of our results with those of the previous studies would be of 
interest in the planning of the future experiments for the charm baryon production 
at the ${\bar P}ANDA$ facility. For this purpose, we chose the ${\bar p}_{lab}$ 
of 15 GeV/$c$. For this beam momentum, results for the total cross section of the 
${\bar p}p \to {\bar \Lambda}_c^- \Lambda_c^+$ reaction have been reported in 
Refs.~\cite{kai94,gor09,kho12}, which are approximately 100, 1.2, and 60 nb, 
respectively. These values are drastically lower than the corresponding 
cross section predicted in our study. On the other hand, our results are in  
close agreement with those of Refs.~\cite{hai10,hai11} even though they have 
given predictions for the cross section only for near-threshold beam momenta. Thus,
even when considering the variation in values predicted in our model due to the 
unconstraint initial and final state distortions and the ansatz of the form factor 
at various vertices, the differences between our cross sections and those of 
Refs.~\cite{kai94,gor09,kho12} are substantial for beam momenta relevant to 
the ${\bar P}ANDA$ experiment.  
 
\begin{figure}[t]
\centering
\includegraphics[width=.50\textwidth]{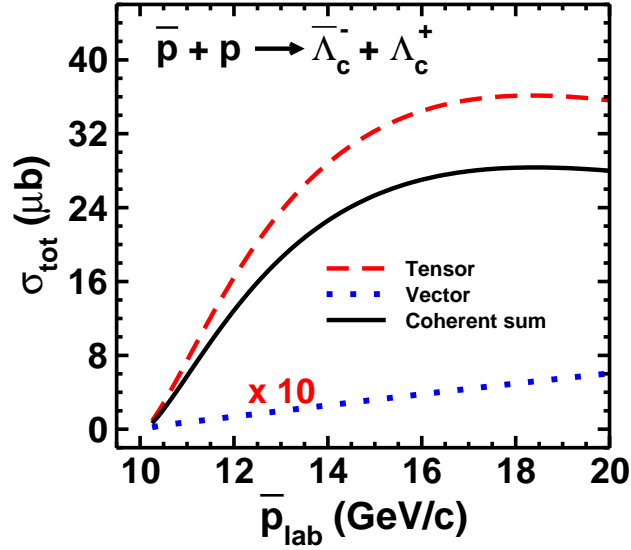}
\caption{(color online)
Contributions of the vector and tensor coupling terms to the $D^{*0}$ exchange 
cross sections for the  ${\bar p}p \to {\bar \Lambda}_c^- \Lambda_c^+$
reaction as a function of the antiproton beam momentum. The contributions of the 
tensor and vector terms are shown by the dashed and dotted lines, respectively.
The solid line represents their coherent sum.
}
\label{fig:Fig5}
\end{figure}

The differential cross section (DCS) provides more valuable information about 
the reaction mechanism. The DCS includes terms that weigh the interference terms 
of various components of the amplitude with angles of the measured outgoing 
particle. Therefore, the structure of the interference terms could highlight the 
contributions of different meson exchanges in different angular regions. In Fig.~6,
we show our results for differential cross sections at the beam momenta of 10.25,
12.25, and 16.25 GeV/$c$. For all three beam momenta, the cross sections are peaked 
in the forward directions. By looking at the relative contributions of the $D^0$ 
and $D^{*0}$ exchange processes that are shown by the dotted and dashed lines, 
respectively, their strongly different characteristics in different angular regions 
become very apparent. While the $D^0$ exchange terms are large in the backward 
directions, those of the $D^{*0}$ exchange are forward peaked. We also note that, 
while there is destructive interference among the two exchange terms at back angles, 
they are strongly constructive in the
forward directions. Even though the individual $D^0$ exchange contributions are   
small at the forward angles, they play an important role in the total cross sections
through the interference terms.

\begin{figure}[t]
\centering
\includegraphics[width=.50\textwidth]{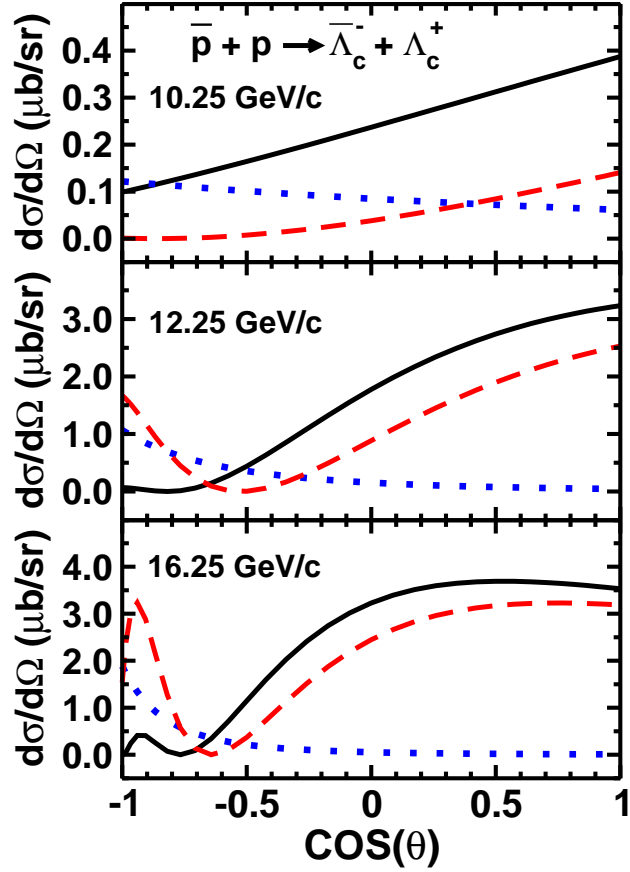}
\caption{(color online)
Differential cross sections for the ${\bar p}p \to {\bar \Lambda}_c^- 
\Lambda_c^+ $ reaction at the beam momenta of 10.25, 12.25, and 
16.25 GeV/$c$ as indicated in the figure. Contributions of the $D^0$  
and $D^{*0}$ exchange processes are shown by dotted and dashed lines, respectively,
in each case. The solid lines represent their coherent sum.
}
\label{fig:Fig6}
\end{figure}

\section{Summary and conclusions}

In summary,  we studied the ${\bar p}p \to {\bar \Lambda}_c^- \Lambda_c^ + $ 
reaction using a phenomenological effective Lagrangian model that involves the
meson-baryon degree of freedom.  The detailed dynamics of the process is 
accounted for by the $t$-channel $D^0$ and $D^{*0}$ exchange diagrams, while 
largely phenomenological initial and final state interactions have been used to 
account for the distortion effects. The coupling constants at various vertices 
have been taken from the $DN$ and ${\bar D}N$ scattering studies reported in 
Refs.~\cite{hai07,hai08,hai11a}. The off-shell corrections at the vertices are 
accounted for by introducing monopole form factors with a cutoff parameter of 
3.0 GeV, which was taken to the same for all the cases. This ansatz for the
form factor and the value of the cutoff parameter $\lambda$ have been checked by 
fitting the data on the $pp \to \Lambda_c^+ X$ reaction measured by the ISR 
Collaboration~\cite{cha87} in Ref.~\cite{hob13}. A further check on the input 
parameters of our model is performed by taking over the same coupling constants 
and the form factor to describe the data on ${\bar p}p \to {\bar \Lambda} \Lambda$ 
reaction under the assumption of SU(4) symmetry. Of course, the value of the cutoff 
parameter cannot be taken over, as the masses of the exchanged mesons for this 
reaction are much smaller. We used a $\lambda$ of 1.7 GeV for this case in line 
with the choice of Refs.~\cite{hai92a,hai92b}. Our  calculations provide 
a good description of the data for this reaction.

The total cross section for the  ${\bar p}p \to {\bar \Lambda}_c^- \Lambda_c^+ $  
vary between 1 and 8 $\mu b$ at near-threshold beam momenta (excess energy between 
1 and 100 MeV). This value agrees with that reported in the coupled-channels 
meson-exchange model calculations of Refs.~\cite{hai10,hai11}. For higher beam 
momenta, the cross sections are larger. They peak around a ${\bar p}_{lab}$ of 
15 GeV/$c$ with a peak value of around 30 $\mu$ b. This value is drastically larger 
than the cross sections for this 
reaction predicted in previous calculations. Since these earlier calculations 
have used different types of models, which by and large invoke the quark degrees of 
freedom in their calculations, it is difficult to locate the reason for the large 
difference between them and our results. This will be understood when the 
${\bar P}ANDA$ experiment performs these measurements once the FAIR facility is 
operational. If the cross sections are as large as predicted in our calculations 
as well as in those of Refs.~\cite{hai10,hai11}, it would be relatively easy to 
measure them at the ${\bar P}ANDA$ experiment.

We noted that the vector meson ($D^{*0}$) exchange terms dominate the cross 
sections for all beam momenta except for those very close to the production 
threshold. The reasons for the large strength of this exchange process are the 
strong tensor coupling of the vector mesons (similar to the large tensor coupling 
of the $\rho$ meson in $NN$ interactions), and the additional momentum dependence 
introduced by the derivative part of the corresponding interaction. Although, 
except for the very-close-to-threshold beam momenta, the individual contributions 
of the $D^0$ exchange terms are relatively weak, they contribute significantly 
through the interference terms.

We found that different meson-exchange processes contribute in different angular
reasons of the differential cross sections, which are generally forward peaked
both at lower as well as higher beam momenta. It was noted that while the $D^0$ 
exchange terms dominate in the backward directions, $D^{*0}$ exchange processes 
are relatively large in the forward angular region. The constructive interference 
between these two exchange processes leads to more forward peaking of the cross 
sections. On the other hand, $D^0$ exchange terms alone yield backward-peaked 
differential cross sections.
 
The initial and final state interactions are the important ingredients of our
model. We treat them within an eikonal approximation-based phenomenological 
method. Generally, the parameters of this model are constrained by fitting to the 
experimental data. Because of the lack of any experimental information, it has not
been possible to test our model thoroughly. Therefore, there may be some 
uncertainty in the absolute magnitudes of our cross sections. Nevertheless, 
we reproduce the data for the ${\bar \Lambda} \Lambda$ production channel, and our 
near-threshold cross sections for the ${\bar \Lambda}_c^- \Lambda_c^+ $ production
are very close to those of Refs.~\cite{hai10,hai11}, where distortion effects have 
been treated more rigorously within a coupled-channels approach. Therefore, the 
large cross sections obtained in our calculations at larger beam momenta as compared 
to those of previous authors are robust and can help in planning of the experiments
to measure this channel at the ${\bar P}ANDA$ facility. 
   
\section{acknowledgments}
 
This work has been supported by the German Research Foundation (DFG) under Grant
No. Le439/8-2 and Helmholtz International Center (HIC) for FAIR and the Council of 
Scientific and Industrial Research (CSIR), India.

\end{document}